\documentclass[letterpaper,english,reprint, aps]{revtex4-2}
\pdfoutput=1

\usepackage[latin9]{inputenc}
\usepackage{color}
\usepackage{amsmath}
\usepackage{amssymb}
\usepackage{graphicx}
\usepackage[normalem]{ulem}

\makeatletter

\pdfpageheight\paperheight
\pdfpagewidth\paperwidth

\providecommand{\tabularnewline}{\\}

\makeatother

\usepackage{babel}
\begin{document}
\title{Calculation of the nonlinear response functions of intra-exciton transitions
in two-dimensional transition metal dichalcogenides}
\author{J. C. G. Henriques$^{1,2}$, H{\o}gni C. Kamban$^{3,4}$, Thomas G. Pedersen{\normalsize{}$^{3,4}$},
N. M. R. Peres$^{1,2}$}
\address{$^{1}$Department and Centre of Physics, and QuantaLab, University
of Minho, Campus of Gualtar, 4710-057, Braga, Portugal}
\address{$^{2}$International Iberian Nanotechnology Laboratory (INL), Av. Mestre
Jos\'e Veiga, 4715-330, Braga, Portugal}
\address{$^{3}$Department of Materials and Production, Aalborg University,
DK-9220 Aalborg {\O}st, Denmark}
\address{$^{4}$Center for Nanostructured Graphene (CNG), DK-9220 Aalborg {\O}st,
Denmark}
\begin{abstract}
In this paper, we study the third-order nonlinear optical response
due to transitions between excitonic levels in two-dimensional transition
metal dichalcogeniedes. To accomplish this, we use methods not applied
to the description of excitons in two-dimensional materials so far
and combined with a variational approach to describe the $1s$ excitonic
state. The aforementioned transitions allow to probe dark states which
are not revealed in absorption experiments. We present general formulas
capable of describing any third-order process. The specific case of
two-photon absorption in WSe\textsubscript{2} is studied. The case
of the circular well is also studied as a benchmark of the theory.
\end{abstract}
\maketitle

\section{Introduction}

Since graphene \citep{novoselov2012roadmap} was first studied the
family of two-dimensional (2D) materials has been expanding, and other
materials such as hexagonal-boron nitride (hBN) \citep{caldwell2019photonics},
phosphorene \citep{carvalho2016phosphorene} and transition metal
dichalcogeniedes (TMDs) \citep{wang2012electronics}, such as MoS\textsubscript{2},
MoSe\textsubscript{2}, WS\textsubscript{2} and WSe\textsubscript{2},
have gained considerable attraction over the years. These last ones
correspond to semiconducting materials with a direct band gap of about
1.5 eV \citep{mak2010atomically}, and are currently extensively studied
due to their remarkable electronic and optical properties.

Like other 2D materials, the optical properties of TMDs are strongly
dependent on their excitonic response \citep{wang2018colloquium}.
When a material is optically excited, if the photon energy is large
enough, electrons may be removed from the valence band to the conduction
band. The electron promoted to the conduction band and the hole left
in the valence band form a quasi-particle due to the Coulomb-like
interaction between them.This particle is similar to a Hydrogen atom,
and it is termed an exciton. Contrary to their 3D counterparts, where
the energy spectrum is well described by a Rydberg series, excitons
in 2D materials present a more complex energy landscape as a consequence
of the nonlocal dielectric screening of the interaction potential
between the electron and the hole \citep{hsu2019dielectric}. Also,
their reduced dimensionality leads to more tightly bound excitonic
states, which are stable even at room temperature \citep{chernikov2014exciton}.

When studying the optical properties of TMDs two distinct regimes
can be identified. The first one corresponds to the case where optical
excitation induces transitions from the excitonic vacuum to a given
state of the exciton, and is termed the excitonic interband regime.
This regime is the origin of the well-known peaks in an absorption
spectrum, corresponding to transitions from the excitonic vacuum to
different $s-$states of the exciton \citep{koperski2017optical}.
In recent years, nonlinear optical effects originated from interband
transitions have been the topic of many works, both experimental and
theoretical. The study of nonlinearities in MoS\textsubscript{2}
was explored in Refs. \citep{wang2014third,soh2018optical,saynatjoki2017ultra,li2013probing},
while Refs. \citep{torres2016third,janisch2014extraordinary} studied
similar effects in WS\textsubscript{2} and Refs. \citep{rosa2018characterization,zeng2013optical}
in WSe\textsubscript{2}. A thorough comparison between the nonlinear
response of different TMDs is presented in Ref. \citep{autere2018optical}.
Studies on the effect of strain and the coupling to exciton-plasmons
have also been performed \citep{liang2020giant,sukharev2018effects}.
In Ref. \citep{taghizadeh2019nonlinear} an analytical study of the
nonlinear optical response of monolayer TMDs was presented. Due to
their broken inversion symmetry TMDs are not centrosymmetric (at least
when stacked in an odd number of layers), and as a consequence both
even and odd orders of non linear optical processes are always permitted
\citep{autere2018nonlinear}. Moreover, these materials shown large
nonlinear optical coefficients \citep{taghizadeh2019nonlinear}, increasing
their potential for applications, such as optical modulators \citep{sun2016optical,wang2015tunable}.
The possibility of characterizing different properties of the 2D material
from their nonlinear optical response has also been considered \citep{karvonen2017rapid,autere2017rapid}.
The second regime one should consider when studying the optical properties
of these systems is associated with transition between the excitonic
energy levels themselves, and we label it as the intra-exciton regime.
This type of excitation can be experimentally realized in a pump-probe
setup, where first the pump laser populates the $1s$ exciton state
and then the probe induces transitions from the $1s$ to the remaining
bound states of the exciton. Recently, in Ref. \citep{pollmann2015resonant},
this type of procedure was implemented to characterize the linear
optical response of WSe\textsubscript{2} in the intra-exciton regime,
and probe the excitonic dark states which are not accessible from
luminescence methods. Contrary to the interband regime, the nonlinear
response associated with optical transitions when the ground state
is already populated remains vastly unstudied. Its comprehension could
unlock new degrees of freedom exploitable in nonlinear optical applications.

Our goal with this paper is to provide a theoretical framework based
on the ideas presented in Refs. \citep{karplus1963variation,hameka1977variational,svendsen1977calculation,svendsen1985variational,svendsen1988variational},
which allows the description of third-order nonlinear optical processes
in the intra-exciton regime, namely the two photon absorption (TPA)
for excitons in WSe\textsubscript{2}. Contrarily to the approach
of a sum over states usually found from time-dependent perturbation
theory, where different excited wave functions are needed, our approach
only requires the $1s$ wave function, which can be accurately described
using variational techniques \citep{pedersen2016exciton,quintela2020colloquium}.
We then expand the perturbed wave function directly in a basis. It follows that, formally, our
approach is equivalent to a sum of states computation approximating excited states
by expanding in the same basis. However, the present approach is conceptully simpler.
The text is organized as follows. In Sec. 2 we present the general
method used to compute the nonlinear third-order optical susceptibility.
This corresponds to a generalization of the approach presented in
Ref. \cite{HenriquesJOSAB} where the linear response was studied. In Sec.
3 we focus on the more interesting problem of excitons in WSe\textsubscript{2},
when the excitonic ground state is already populated and the optical
excitation induces transitions between the excitonic levels. A section
with our final remarks and an appendix close the paper.
\begin{figure}[h]
\centering{}\includegraphics[width=8.5cm]{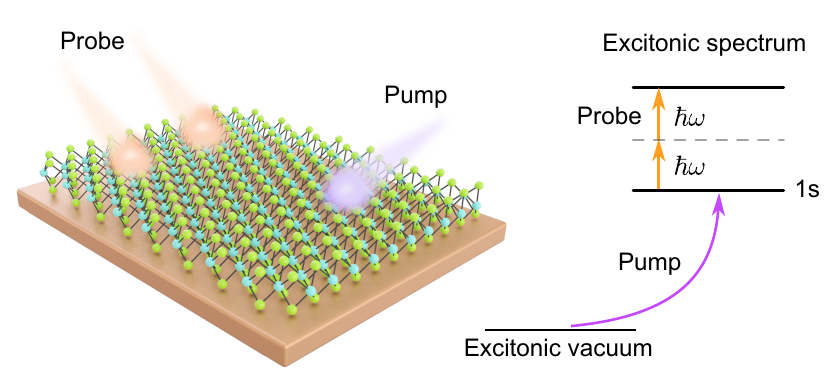}\caption{Schematic representation of the two photon absorption process in WSe\protect\textsubscript{2}
excitons when the 1$s$ excitonic state is already populated.}
\end{figure}

\section{Nonlinear third-order optical response}

In the first part of this section we will give a detailed description
of a method to compute the third-order optical susceptibility of a
given system. The only requirement is that the ground-state wave function
of the system is known (at least approximately). This method contrasts
with the usual sum over states where both the ground state and the
excited-state wave functions are needed. The presented approach is
based on Refs. \citep{karplus1963variation,hameka1977variational,svendsen1977calculation,svendsen1985variational,svendsen1988variational}
and corresponds to an extension of what was recently used in Ref.
\cite{HenriquesJOSAB}
regarding the linear optical response. In the second part of the section
the problem of a circular potential well will be studied as a first
application of the formalism. This example will set the stage for
the posterior study of two\textendash dimensional excitons in WSe\textsubscript{2}.

\subsection{Outline of the Method}

\subsubsection{Third-order susceptibility}

Since we will be interested in computing the third-order nonlinear
response, we start by introducing the expression for the third-order
optical susceptibility, as derived from perturbation theory. Throughout
the work we will use atomic units unless stated otherwise. Following
Ref. \citep{orr1971perturbation} we write the third-order susceptibility
as
\begin{align}
 & \chi_{\alpha\beta\gamma\delta}^{(3)}({\omega_{\sigma}};\omega_{1},\omega_{2},\omega_{3})=\nonumber \\
= & \frac{1}{3!}\mathcal{P}\Bigg\{\sum_{n,m,l\neq0}\frac{\langle0|\mathbf{d}_{\alpha}|n\rangle\langle n|\mathbf{d}_{\beta}|m\rangle\langle m|\mathbf{d}_{\gamma}|l\rangle\langle l|\mathbf{d}_{\delta}|0\rangle}{(E_{n0}-\omega_{\sigma})(E_{m0}-\omega_{2}-\omega_{3})(E_{l0}-\omega_{3})}\nonumber \\
- & \sum_{n,m\neq0}\frac{\langle0|\mathbf{d}_{\alpha}|n\rangle\langle n|\mathbf{d}_{\beta}|0\rangle\langle0|\mathbf{d}_{\gamma}|m\rangle\langle m|\mathbf{d}_{\delta}|0\rangle}{(E_{n0}-\omega_{\sigma})(E_{m0}-\omega_{2})(E_{m0}+\omega_{1})}\Bigg\},\label{eqOrr chi3}
\end{align}
where, $E_{n0}=E_{n}-E_{0}$ is the energy difference between the
levels $\vert n\rangle$ and $\vert0\rangle$, $\mathbf{d}$ is the
dipole moment, $\{\alpha,\beta,\gamma,\delta\}$ are indexes corresponding
to different spatial orientations ($x$ or $y$), $\omega_{\sigma}=\omega_{1}+\omega_{2}+\omega_{3}$,
$\mathcal{P}$ is the permutation operator of the pairs $(\alpha,-\omega_{\sigma};\beta,\omega_{1};\gamma,\omega_{2};\delta,\omega_{3})$
and $|n\rangle$ corresponds to the unperturbed states of the system,
with $|0\rangle$ its ground state. The direct application of Eq.
(\ref{eqOrr chi3}) corresponds to the sum over states approach.
Since the different sums run over all the excited states of the system,
this way of calculating the optical susceptibility presents the major
drawback of requiring the knowledge of all the excited-state wave
functions, at least in a naive approach (obviously one can expand
the unknown eigenstates in a complete basis and obtain the expansion
coefficients. This latter approach can be seen as an alternative to
the method developed in the Appendix \ref{secComputing-the-new}).
Although in simple systems the exact wave functions may be trivially
known, in more complex ones they may be elusive (this is precisely
the case of excitons in 2D materials to be discussed ahead).

In order to avoid the usual sum over states method we follow the ideas
of Refs. \citep{karplus1963variation,hameka1977variational,svendsen1977calculation,svendsen1985variational,svendsen1988variational}.
Doing so, we write the time-dependent Schr\"{o}dinger equation
\begin{equation}
\left[H_{0}+\mathbf{d}\cdot\boldsymbol{\mathcal{E}}(t)\right]|\psi(t)\rangle=i\frac{\partial}{\partial t}|\psi(t)\rangle.
\end{equation}
where $H_{0}$ corresponds to the unperturbed Hamiltonian of a given
system (this may contain a kinetic and a potential term), $\mathbf{d}\cdot\boldsymbol{\mathcal{E}}(t)$
describes the interaction of the system with an external time-dependent
harmonic electric field $\boldsymbol{\mathcal{E}}(t)$ in the dipole
approximation, and $|\psi(t)\rangle$ is the state vector of the system
in the presence of the external electric field. Next, we expand $|\psi(t)\rangle$
in powers of $\mathcal{E}$ as
\begin{align}
|\psi\rangle & =|0\rangle e^{-iE_{0}t}+\mathcal{E}_{\alpha}|\psi_{\alpha}\rangle e^{-i(E_{0}-\omega_{a})t}\nonumber \\
 & +\mathcal{E}_{\alpha}\mathcal{E}_{\beta}|\xi_{\alpha\beta}\rangle e^{-i(E_{0}-\omega_{a}-\omega_{b})t}+...
\end{align}
where $\boldsymbol{\mathcal{E}}_{\alpha}e^{i\omega_{a}t}$ refers
to an harmonic electric field applied along the $\alpha$ direction
(either $x$ or $y$) with frequency $\omega_{a}$, $E_{0}$ is the
energy of the unperturbed ground state of the system and $|\psi_{\alpha}\rangle$
and $|\xi_{\alpha\beta}\rangle$ are yet to be determined. Inserting
this in the time-dependent Schr\"{o}dinger equation and grouping equivalent
terms in $\mathcal{E}$, up to second order in the electric field,
we find the following three equations
\begin{align}
0 & =\left(H_{0}-E_{0}\right)|0\rangle\\
0 & =\left(H_{0}-E_{0}+\omega_{a}\right)|\psi_{\alpha}\rangle+\mathbf{d}_{\alpha}|0\rangle\label{eqpsi_alpha_def}\\
0 & =\left(H_{0}-E_{0}+\omega_{a}+\omega_{b}\right)|\xi_{\alpha\beta}(\omega_{a},\omega_{b})\rangle+\mathbf{d}_{\beta}|\psi_{\alpha}(\omega_{a})\rangle.\label{eqxi_alpha_beta_def}
\end{align}
The first one simply states the eigenvalue relation for the ground
state of the system in the absence of the external electric field.
The second and third ones define the $|\psi_{\alpha}\rangle$ and
$|\xi_{\alpha\beta}\rangle$, respectively. Expanding these two states
in the basis of the eigenstates of $H_{0}$ one easily arrives at
\begin{align}
|\psi_{\alpha}(\omega_{a})\rangle & =-\sum_{n\neq0}\frac{\langle n|\mathbf{d}_{\alpha}|0\rangle}{E_{n}-E_{0}+\omega_{a}}|n\rangle,\label{eqpsi_alpha}\\
|\xi_{\alpha\beta}(\omega_{a},\omega_{b})\rangle & =-\sum_{n\neq0}\frac{\langle n|\mathbf{d}_{\beta}|\psi_{\alpha}(\omega_{a})\rangle}{E_{n}-E_{0}+\omega_{a}+\omega_{b}}|n\rangle,\label{eqxi_alpha_beta}
\end{align}
where we assumed $\langle0|\psi_{\alpha}\rangle=0$ and $\langle0|\xi_{\alpha\beta}\rangle=0$.
The first requirement corresponds to choosing a coordinate system placing $\langle0|\mathbf{d}|0\rangle$ at the origin, which is always possible. The second assumption will be discussed further ahead. Now we note
that with the introduction of $|\psi_{\alpha}\rangle$ and $|\xi_{\alpha\beta}\rangle$
we are able to rewrite Eq. (\ref{eqOrr chi3}) without the sums over
the excited states, hence
\begin{align}
\chi_{\alpha\beta\gamma\delta}^{(3)} & =\frac{1}{3!}\mathcal{P}\Big\{-\langle\psi_{\alpha}\left(-\omega_{\sigma}^{*}\right)|\mathbf{d}_{\beta}|\xi_{\delta\gamma}(-\omega_{3},-\omega_{2})\rangle\nonumber \\
 & +\langle0|\mathbf{d}_{\alpha}|\psi_{\beta}(-\omega_{\sigma})\rangle\langle\psi_{\gamma}\left(-\omega_{2}^{*}\right)|\psi_{\delta}(\omega_{1})\rangle\Big\}.\label{eqchi3 Karplus}
\end{align}
Thus, using the ideas of Ref. \citep{karplus1963variation,hameka1977variational,svendsen1977calculation,svendsen1985variational,svendsen1988variational},
we shifted the problem away from the sum over states, to the determination
of two new state vectors $|\psi_{\alpha}\rangle$ and $|\xi_{\alpha\beta}\rangle$.
After being determined, these state vectors allow us to access the
third-order susceptibility through the computation of only three matrix
elements. Note that we are considering the possibility of the frequencies
to be complex valued. We do so in order to obtain both the real and
imaginary parts of $\chi^{(3)}$ . This is achieved by shifting the
energies by a small imaginary part, that is $\omega\rightarrow\omega+i\delta$.

\subsubsection{Computing the new state vectors}

Now that an alternative path to the sum over states was found, we
are left with the task of determining $|\psi_{\alpha}\rangle$ and
$|\xi_{\alpha\beta}\rangle$. Computing these quantities using Eq.
(\ref{eqpsi_alpha}) and (\ref{eqxi_alpha_beta}) would reverse
our progress, and leave us again with a problem requiring the calculation
of a sum over states. To continue with the calculations, we follow
Ref. \citep{karplus1963variation} and introduce the functionals
\begin{align}
\mathcal{J} & =\langle\psi_{\alpha}(\omega)|H_{0}-E_{0}+\omega|\psi_{\alpha}(\omega)\rangle\nonumber \\
 & +\langle\psi_{\alpha}(\omega)|\mathbf{d}_{\alpha}|0\rangle+\langle0|\mathbf{d}_{\alpha}|\psi_{\alpha}(\omega)\rangle,\label{eqJ_fiunctional}
\end{align}
and
\begin{align}
\mathcal{K} & =\langle\xi_{\alpha\beta}(\omega_{a},\omega_{b})|\left(H_{0}-E_{0}+\omega_{a}+\omega_{b}\right)|\xi_{\alpha\beta}(\omega_{a},\omega_{b})\rangle\nonumber \\
 & +\langle\xi_{\alpha\beta}(\omega_{a},\omega_{b})|\mathbf{d}_{\beta}|\psi_{\alpha}(\omega_{a})\rangle+\langle\psi_{\alpha}(\omega_{a})|\mathbf{d}_{\beta}|\xi_{\alpha\beta}(\omega_{a},\omega_{b})\rangle.\label{eqK_functional}
\end{align}
Minimizing $\mathcal{J}$ with respect to $|\psi_{\alpha}\rangle$
and $\mathcal{K}$ with respect to $|\xi_{\alpha\beta}\rangle$ allows
us to explicitly compute these new state vectors. . Moreover, we note
that the minimization of these functionals is equivalent to directly
solving Eq. (\ref{eqpsi_alpha_def}) and (\ref{eqxi_alpha_beta_def}),
where $|\psi_{\alpha}\rangle$ and $|\xi_{\alpha\beta}\rangle$ were
first introduced.

Since we will be interested in 2D systems, the first step in our procedure
is to confine our system within a disk of radius $R$. If the problem
we are interested in is not naturally bounded, we can first force
it to be defined inside a disk of finite radius, and later chose $R\gg1$
and check the convergence of the results by varying $R$. This procedure
is always possible as long as the wave functions vanish for a large
enough distance away from the origin. After this is done we can expand
$|\psi_{\alpha}\rangle$ and $|\xi_{\alpha\beta}\rangle$ in a Fourier-Bessel
series with a normalised radial basis
\begin{equation}
j_{ln}(r)=\frac{\sqrt{2}J_{l}\left(\frac{z_{ln}r}{R}\right)}{J_{l+1}(z_{ln}) R}
,\label{Bessel_basis}
\end{equation}
where $J_{l}(z)$ is the Bessel function of the first kind of
$l$'th order, $z_{ln}$ corresponds to the $n$'th zero of $J_{l}(z)$, and
$R$ is the radius of the disk where the problem is defined. In terms of this basis,
\begin{align}
\psi_{\alpha}(\omega;\mathbf{r}) & =\frac{1}{\sqrt{2\pi}}\sum_{n=1}^{N}\sum_{l=\pm1}c_{ln}^{\alpha}(\omega)j_{ln}(r)e^{il\theta}\nonumber \\
 & =\frac{1}{\sqrt{2\pi}}\sum_{n=1}^{N}\left[c_{+,n}^{\alpha}(\omega)e^{i\theta}-c_{-,n}^{\alpha}(\omega)e^{-i\theta}\right]j_{1n}(r),\label{eqpsi_alpha_Bessel}
\end{align}
and
\begin{equation}
\xi_{\alpha\beta}(\omega_{a},\omega_{b};\mathbf{r})=\frac{1}{\sqrt{2\pi}}\sum_{n=1}^{N}\sum_{l=-\infty}^{\infty}\zeta_{ln}^{\alpha\beta}(\omega_{a},\omega_{b})j_{ln}(r)e^{il\theta},\label{eqxi_alpha_beta_Bessel}
\end{equation}
where $N$
is the number of functions in the radial basis, $c_{ln}^{\alpha}$ and $\zeta_{ln}^{\alpha\beta}$
are the expansion coefficients, and $(r,\theta)$ are polar coordinates.
Although we choose to work with a Fourier-Bessel basis, other options
could have been used, e.g. orthogonal polynomials or Sturmian functions.
Now, we insert these expressions in the definitions of $\mathcal{J}$
and $\mathcal{K}$ and minimize each functional with respect to the
$c_{ln}^{\alpha}$ and $\zeta_{ln}^{\alpha\beta}$ respectively. Doing
so we arrive at two linear system of equations whose solutions define
the expansion coefficients. In Appendix \ref{secComputing-the-new}
we give the detailed description of the necessary steps to obtain
the linear system of equations, which is numerically well behaved
and can be easily solved. Also discussed in the Appendix is the implication
of the condition $\langle0|\xi_{\alpha\beta}\rangle=0$, which imposes
a restriction on the coefficient $\zeta_{01}^{\alpha\beta}$, requiring
special care when dealing with the term $l=0$ in the functional $\mathcal{K}$.

\subsection{The case of the circular well}

Up to this point have introduced the third-order optical susceptibility,
and presented a way of computing it without a sum over states. To
do this, two new state vectors, $|\psi_{\alpha}\rangle$ and $|\xi_{\alpha\beta}\rangle$,
were introduced. The necessary steps to find both $|\psi_{\alpha}\rangle$
and $|\xi_{\alpha\beta}\rangle$ were already briefly discussed and
their detailed description is given in Appendix \ref{secComputing-the-new}.
Now, as a first application of the ideas presented so far, we will
study the problem of a circular well. This studywill allow for a concrete
application of the general expressions previously derived, as well
as gaining some intuition that will prove helpful when the excitonic
problem is studied ahead.

Consider a particle with mass $\mu$ trapped inside a circular well
of radius $R$. The Hamiltonian of such a system reads
\begin{equation}
H=-\frac{1}{2\mu}\nabla^{2},\quad0\leq r/R<1.
\end{equation}
The eigenstates are given by
\begin{equation}
\psi_{nm}(r,\theta)=\frac{1}{\sqrt{2\pi}}j_{mn}(r)e^{im\theta}.
\end{equation}
Following common practice, we label $n$
as the principal quantum number and $m$ as the angular quantum number.
The energy spectrum reads
\begin{equation}
E_{nm}=\frac{1}{2\mu}\left(\frac{z_{mn}}{R}\right)^{2}.\label{eqcircular well energy}
\end{equation}
The ground-state wave function is $\psi_{\rm GS}(r)=j_{01}(r)/\sqrt{2\pi}$.

\begin{figure}[h]
\centering{}\includegraphics[width=8.5cm]{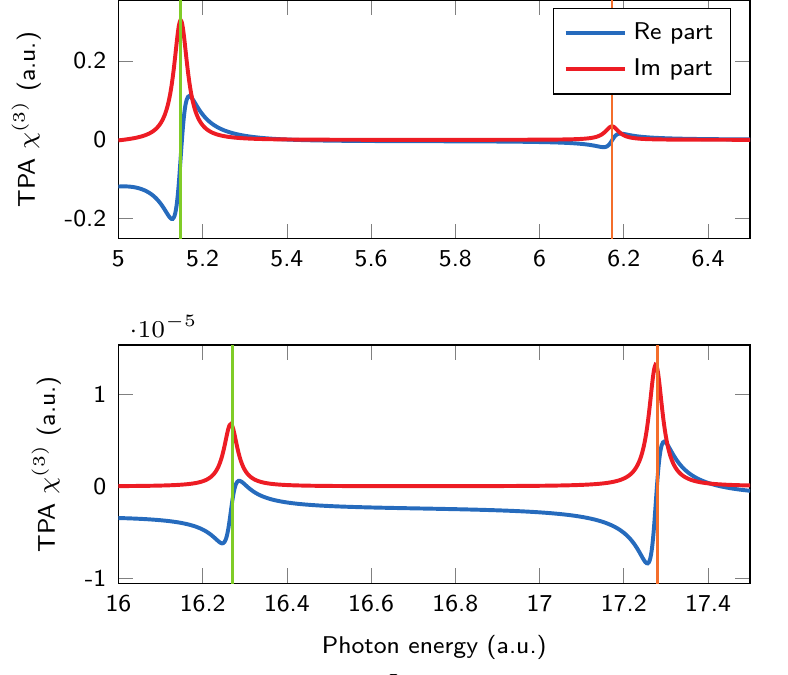}\caption{\label{figDisk_TPA}Plot of the two-photon absorption (TPA) third-order
susceptibility as a function of the photon energy for a particle with
mass $\mu=1$ in a circular well of radius $R=1$. Both quantities
are presented in atomic units (a.u.). The resonances marked with the
orange lines correspond to transitions from the ground state (1$s$)
to the states 2$s$ and $3s$ with the absorption of two photons.
The resonances marked with the green lines are associated with the
transitions $1s\rightarrow1d$ and $1s\rightarrow2d$. As the principal
quantum number of the final state increases, the oscillator strength
of the resonance decreases. In order to obtain the real and imaginary
parts we introduced a small shift in the frequency $\omega$, that
is, $\omega\rightarrow\omega+i\delta$ with $\delta=0.02$ a.u.. $N=5$
basis functions were used.}
\end{figure}

There are many nonlinear third-order optical processes \citep{boyd2020nonlinear}.
To be definitive, let us now focus on a specific third-order nonlinear
optical process. We will be interested in computing the $xxxx$ component
of the two photon absorption (TPA) third-order susceptibility $\chi_{xxxx}^{{\rm TPA}}(\omega)=\chi_{xxxx}^{(3)}(-\omega;\omega,-\omega,\omega)$.
Using Eq. (\ref{eqchi3 Karplus}), we write 
\begin{align}
\chi_{xxxx}^{{\rm TPA}} & =\frac{1}{3!}\mathcal{P}\Big\{-\langle\psi_{x}\left(-\omega^{*}\right)|\mathbf{d}_{x}|\xi_{xx}(-\omega,\omega)\rangle\nonumber \\
 & +\langle0|\mathbf{d}_{x}|\psi_{x}(-\omega)\rangle\langle\psi_{x}\left(-\omega^{*}\right)|\psi_{x}(\omega)\rangle\Big\}.\label{eqchi_TPA}
\end{align}
To obtain the TPA spectrum we need only compute $\psi_{x}$ and $\xi_{xx}$
using the procedure discussed previously, and then evaluate three
matrix elements. A detailed description of this is given in Appendix
\ref{secDetails-on-the}. The fact that the potential vanishes inside
the disk of radius $R$ makes this a simple application of the ideas
so far introduced. Considering $\mu=1$ and $R=1$, using $N=5$ basis
functions, and accounting for all the necessary permutations in Eq.
(\ref{eqchi_TPA}), we obtain the results depicted in Fig. \ref{figDisk_TPA}.
This value of $N$ already allows the results to converge; increasing
it produces no change in the TPA spectrum. In order to obtain the
real and imaginary parts of $\chi_{xxxx}^{{\rm TPA}}(\omega)$ we
introduced a small imaginary shift in the frequency $\omega$, i.e.
$\omega\rightarrow\omega+i\delta$. The resonances that appear in
Fig. \ref{figDisk_TPA} have two distinct origins the ones marked
with the orange lines correspond to transitions from the ground state
(which we call the 1$s$ state) to other $s-$states (where the angular
quantum number is $m=0$) with the absorption of two photons; the
ones marked with the green lines are associated with transitions from
the ground state to $d-$states ($m=2$), due to the absorption of
two photons. As the principal quantum number of the final state increases,
the oscillator strength of the transition decreases and the resonances
become less pronounced. One of the main advantages of studying the
circular well lies in its parabolic energy spectrum (see Eq. (\ref{eqcircular well energy})),
since the energy levels are significantly separated, allowing for
an effortless identification of the relevant optical transitions.

\section{Two-photon absorption for excitons in ${\rm WS}$${\rm e}_{2}$\protect\textsubscript{}}

In the previous section we presented a method to compute the third-order
optical susceptibility of a system without performing a sum over states.
Afterwards we explored the problem of a circular disk as a first application
of the formalism. Now, in the current section, we will discuss the
more interesting topic of 2D excitons in WSe\textsubscript{2}. More
accurately, we will study the third-order optical response associated
with transitions from the ground state ($1s$) to excited states of
the 2D exciton. This problem is the natural extension of the work
done in \cite{HenriquesJOSAB} and the computed physical quantity can be
measured experimentally in a pump-probe experiment.

The Hamiltonian that describes the excitonic problem reads
\begin{equation}
H_{0}=-\frac{1}{2\mu}\nabla^{2}+V_{{\rm RK}}(r),\label{eqHamiltonian Exciton}
\end{equation}
where $\mu$ is the reduced mass of the electron\textendash hole pair,
$\nabla^{2}$ is the 2D Laplacian and $V_{{\rm RK}}(r)$ is the Rytova-Keldysh
potential \citep{rytova1967,keldysh1979coulomb}
\begin{equation}
V_{{\rm RK}}=-\frac{\pi}{2r_{0}}\left[\mathbf{H}_{0}\left(\frac{\kappa r}{r_{0}}\right)-Y_{0}\left(\frac{\kappa r}{r_{0}}\right)\right],\label{eqRKPot}
\end{equation}
where $\kappa$ is the mean dielectric constant of the media above
and below the TMD, $r_{0}$ is an intrinsic parameter of the 2D material
which can be interpreted as an in-plane screening length and is related
to the effective thickness of the material; $\mathbf{H}_{0}$ and
$Y_{0}$ are the Struve function and the Bessel function of the second
kind, both of order 0, respectively. This potential is the solution
of the Poisson equation for a charge embedded in a thin film. For
large distances the Rytova-Keldysh presents a Coulomb$-1/\kappa r$
tail, but diverges logarithmically near the origin.

\begin{figure}[h]
\centering{}\includegraphics[width=8.5cm]{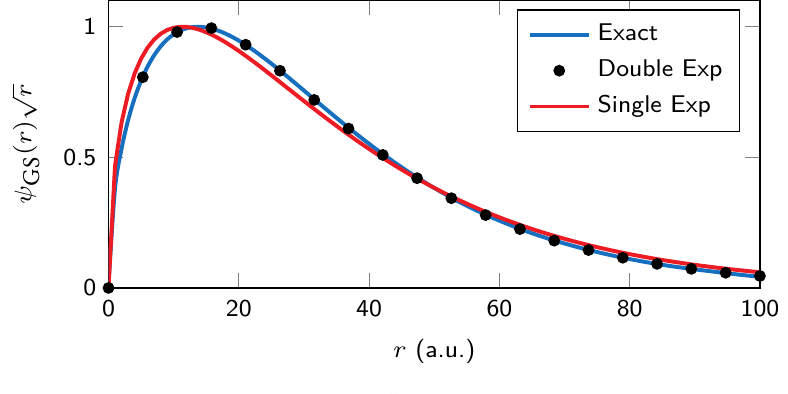}\caption{\label{figWF_comparison}Comparison between the excitonic ground-state
wave function obtained exactly (using a shooting algorithm), and the
ones obtained with the single and double variational ansaetze. Although
the single exponential approach already produces a good result, the
one obtained with the double exponential is clearly superior. The
values of the variational parameters were obtained from the minimization
of the expected value of the Hamiltonian of Eq. (\ref{eqRKPot}).
The values of Table \ref{tabParameters} were used.}
\end{figure}

While in the previous section we showed the usefulness of our approach
when we computed $\chi_{xxxx}^{{\rm TPA}}(\omega)$ for the circular
well without evaluating a sum over states, the true potential of the
method is clearer when it is applied to the excitonic problem. Contrary
to the circular well, or even the Hydrogen atom, the 2D excitonic
problem does not offer a simple analytical solution. In fact, computing
the wave functions of the different excitonic states is an involved
problem, where the wave functions are only known either numerically
or semi-analytically (where the wave functions can be computed analytically
up to a set of numerical coefficients). In the present approach, perturbed wave functions
are computed directly by expanding in a basis without the intermediate step of
finding excited states. We have shown that in order to apply the formalism
presented in Sec. 2 only the wave function of the exciton ground state
is required. Finding this wave function is a considerably simpler
task, and in order to work with an analytical expression we follow
a variational approach. To obtain accurate results for the optical
susceptibility, it is necessary to use an appropriate ground-state
wave function. It is thus imperative that our variational ansatz produces
an excellent description of the exact solution. A first proposal for
the variational ansatz, inspired by the 2D Hydrogen atom, could be
a single exponential such as $\exp(-ar)$, where $a$ is a variational
parameter. Although this already produces a good description of the
exact ground-state wave function, we turn to Ref. \citep{pedersen2016exciton},
where a more sophisticated double exponential ansatz was proposed
\begin{equation}
\psi_{{\rm GS}}(r)=\frac{1}{\sqrt{\mathcal{N}}}\left(e^{-ar}+be^{-a\gamma r}\right),\label{eqvariational ansatz}
\end{equation}
with $a$, $b$ and $\gamma$ variational parameters and $\mathcal{N}$
a normalization constant. As one can observe in Fig. \ref{figWF_comparison},
where the exact wave function is compared with the single and double
exponential ansaetze for excitons in WSe\textsubscript{2}, Eq. (\ref{eqvariational ansatz})
produces an outstanding description of the exact solution, the latter
computed with a numerical shooting algorithm. As we have already noted in Sec. 2, the domain of our problem should
be enclosed within a disk of finite radius. However, the excitonic
problem is usually considered as an unbounded one, and the wave functions
extend up to infinity, where they smoothly vanish. In practice, we find that the ground state wave function
has vanished at $r=R$ to an excellent approximation and, hence, contributions from the edge are negligible.

As in the case of the circular well, let us consider the $xxxx$ component
of the TPA susceptibility associated with transitions from the 1$s$
to the excited excitonic states. Its general expression was already
given in Eq. (\ref{eqchi_TPA}). To evaluate the TPA spectrum we
return once more to the problem of minimizing the $\mathcal{J}$ and
$\mathcal{K}$ functionals.. A difference relatively to the circular
well lies in the value of $\zeta_{01}$. The orthogonality of Bessel
functions on a disk implied that $\zeta_{01}=0$ for the circular
well. For the excitonic problem no simple rule applies, and the value
of $\zeta_{01}$ must be determined from Eq. (\ref{eqzeta_01}).

\begin{figure}[h]
\centering{}\includegraphics[width=8.4cm]{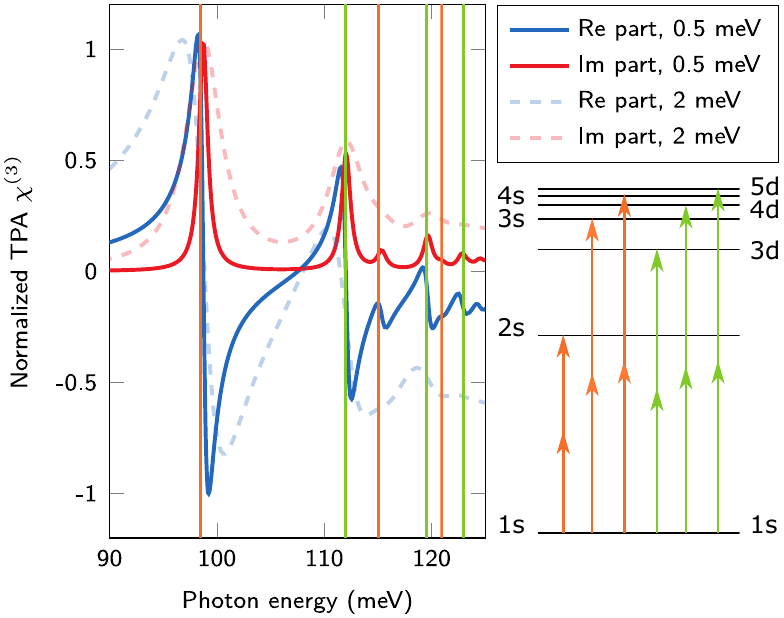}\caption{\label{figExciton spectrum}Real and imaginary parts of the TPA susceptibility
(normalized to its maximum valued) for two different degrees of disorder
(0.5 and 2 meV). The resonances correspond to transitions from the
$1s$ to the $2s$, $3s$ and $4s$ states (marked in orange) and
to the $3d$, $4d$ and $5d$ states (marked in green) with the absorption
of two photons. The energies of these transitions were computed from
the binding energies of the involved states which in turn were obtained
numerically using a shooting algorithm. A diagram of the optical transitions
behind the resonances is also depicted.}
\end{figure}

Using the parameters given in Table \ref{tabParameters} the TPA
spectrum plotted in Fig. \ref{figExciton spectrum} was obtained.
\begin{table}[h]
\centering{}%
\begin{tabular}{|c|c|c|c|c|}
\hline 
$\mu$ & $\kappa$ & $r_{0}$ & $R$ & $N$\tabularnewline
\hline 
\hline 
0.167 & 3.32 & 51.9 & 2500 & 150\tabularnewline
\hline 
\end{tabular}\caption{\label{tabParameters}Parameters used to compute the TPA spectrum
due to intra-excitonic transition in WSe\protect\textsubscript{2}.
All the quantities are given in atomic units. The values of $\mu$,
$\kappa$ and $r_{0}$ were taken from Ref. \citep{pollmann2015resonant}.
The value of $R$ was chosen in order to have $\psi_{{\rm GS}}(R)\approx 0$.
The value of $N$ allowed the results to converge.}
\end{table}
 The value of $R$ was chosen such that $\psi_{{\rm GS}}(R)\approx 0$.
A small value for the radius modifies the results due to the effect
of the confinement that we introduced in the problem. A larger value
for $R$ suppresses the effect of the confinement at the cost of increased
convergence difficulty, as a larger $R$ requires a higher $N$. We
found that $R=2500$ allows an accurate description of the excitonic
problem, while keeping the method efficient. The number of functions
that make up the Fourier-Bessel basis was chosen as the minimum $N$
which when increased leaves the result unchanged. The results proved
to be stable with respect to small variations of both $N$ and $R$,
and inspection of the different coefficients that appear in the Fourier-Bessel
expansions confirmed the convergence of the results.
Looking at Fig. \ref{figExciton spectrum} we observe a similar result
to the one found for the circular well. In order to identify the optical
transitions behind each resonance we computed the energy of the different
excitonic states using a shooting algorithm, and from there the energies
of the transitions from the $1s$ to other states were computed. This
allowed us to assert that the resonances in Fig. \ref{figExciton spectrum}
are due to transitions from the ground state ($1s$) to the $2s$,
$3s$ and $4s$ states (marked in orange) and to the $3d$, $4d$
and $5d$ states (marked in green) with the absorption of two photons.
The identification of the optical transitions behind each resonance
was also facilitated by the intuition gained from the study of the
circular well. The real and imaginary parts of the TPA susceptibility
were obtained by introducing a small imaginary part on the photon
energy $\omega\rightarrow\omega+i\delta$, where the parameter $\delta$
characterizes the broadening of the excitonic level. As expected,
increasing the value of $\delta$ leads to broader and less intense
resonances. For large values of $\delta$ a nonphysical shift of the
resonance starts to appear. This effect is the main limitation of
our approach. Currently it is possible to study this kind of system
with a linewidth of about 20 meV for a sample on glass \citep{koirala2016homogeneous},
and for encapsulated systems at low temperatures spectral broadening
as low as 2 meV can be achieved \citep{robert2018optical}. From the
results depicted in Fig. \ref{figExciton spectrum}, where the maximum
broadening is 2 meV, we expect that experimental measurements of the
TPA performed on encapsulated systems should be able to clearly capture
the resonances originated by the $1s\rightarrow2s$, $1s\rightarrow3d$
and $1s\rightarrow3s$ transitions. In order to capture more resonances
it is necessary to decrease the linewidth, or change the studied material
to another where the excitonic resonances are further apart (such
as hBN).

\begin{figure}[h]
\centering{}\includegraphics[width=8cm]{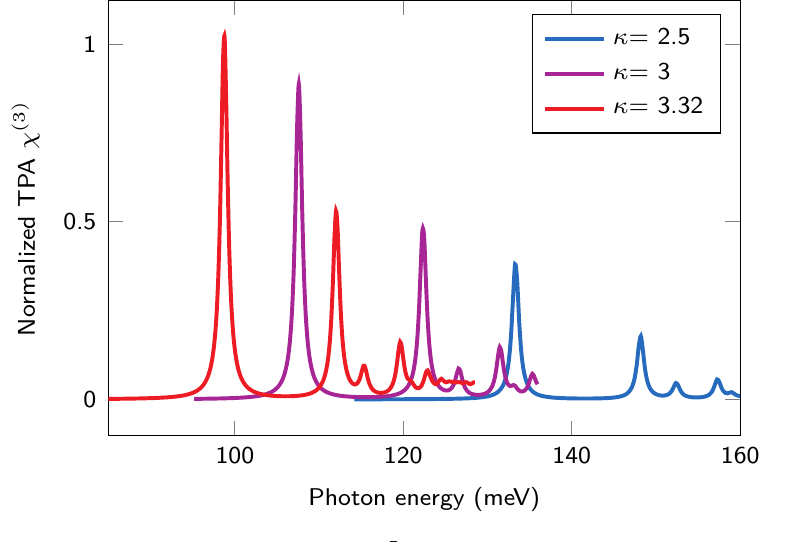}\caption{\label{figDielectric comparison}Comparison of the TPA spectrum for
three different dielectric environments, $\kappa=3.32$, $\kappa=3$
and $\kappa=2.5$. As the dielectric constant decreases the excitons
become more tightly bound, the energetic difference between the ground
state and the excited states increases and the resonances are shifted
to higher energies.}
\end{figure}

In Fig. \ref{figDielectric comparison} we study the role of the
dielectric environment on the TPA spectrum. This parameter appears
in the calculation inside the Rytova-Keldysh potential. As the dielectric
screening is reduced, that is $\kappa$ decreases, the excitons become
more tightly bound. As a consequence, the energy difference between
the ground state and the excited excitonic states increases. This
behavior is reflected in Fig. \ref{figDielectric comparison},
where we observe a blue-shift of the resonances as $\kappa$ decreases.
Moreover, we also observe that the oscillators strength decreases
with decreasing dielectric screening.

\section{Conclusion}

In this work, following the ideas of Refs. \citep{karplus1963variation,hameka1977variational,svendsen1977calculation,svendsen1985variational,svendsen1988variational},
we developed a method to study nonlinear third order processes involving
transitions from the 1$s$ to excited excitonic states. The usual
approach to this type of problem would require the knowledge of all
the excited states in order to compute the different matrix elements
that appear in Eq. (\ref{eqOrr chi3}). The excited states wave functions
are often computed by expanding them in a given basis, e.g. Bessel-Fourier,
followed by the diagonalization of the Hamiltonian. This yields the
sets of coefficients that define the wave functions of the different
excited states, which can then be used to evaluate the many matrix
elements in the sum over states. At odds with this procedure, our
approach avoids the sum over states, and requires only three wave
functions: the ground state wave function, which can be described
using a variational ansatz with high accuracy (see Fig. \ref{figWF_comparison});
and two wave functions defined by Eqs. (\ref{eqpsi_alpha}) and (\ref{eqxi_alpha_beta})
which we determined through an expansion in a Bessel-Fourier basis.

The main result of our work is the TPA spectrum which presents resonances
associated with transitions from the $1s$ state to the remaining
$s-$states as well as from the $1s$ to the $d-$states with the
absorption of two photons. In high purity systems different resonances
should be resolvable. However, in systems with a significant spectral
broadening only the $1s\rightarrow2s$ resonance should be identifiable.
When the role of dielectric screening was studied, a blue shift of
the resonances was observed with decreasing dielectric constant, in
agreement with the increased exciton binding energy and higher energy
separation between the ground state and the excited states. We focused
on the case of excitons in WSe\textsubscript{2}, but other materials
may easily be explored using the method.

Although we focused primarily on the $xxxx$ component of the TPA
susceptibility, we presented general formulas capable of describing
any third-order process in an arbitrary system as long as its ground
state wave function is know, either exactly or approximately. The
study of systems where confinement plays a significant role, such
as quantum dots, may also be treated with this approach.

\section*{Acknowledgements}

N.M.R.P acknowledges support by the Portuguese Foundation for Science
and Technology (FCT) in the framework of the Strategic Funding UIDB/04650/2020.
J.C.G.H. acknowledges the Center of Physics for a grant funded by
the UIDB/04650/2020 strategic project.  N.M.R.P. acknowledges support
from the European Commission through the project ``Graphene-Driven
Revolutions in ICT and Beyond'' (Ref. No. 881603, CORE 3), COMPETE
2020, PORTUGAL 2020, FEDER and the FCT through projects POCI-01-0145-FEDER-028114, POCI-01-0145-FEDER-028887, 
PTDC/NAN-OPT/29265/2017.  H.C.K. and T.G.P. gratefully acknowledge
financial support by the Center for Nanostructured Graphene (CNG),
which is sponsored by the Danish National Research Foundation, Project
No. DNRF103. 

\appendix

\section{Computing the new state vectors\label{secComputing-the-new}}

In this appendix we will give a detailed description of how to obtain
the linear systems whose solution defines the state vectors $|\psi_{\alpha}\rangle$
and $|\xi_{\alpha\beta}\rangle$. Let us consider the $H_{0}$ to
be the unperturbed Hamiltonian of a given system which in general
can be written as
\begin{equation}
H_{0}=-\frac{1}{2\mu}\nabla^{2}+V(r),\label{eqH_0 general}
\end{equation}
where the first term, with $\mu$ a mass term and $\nabla^{2}$ the
2D laplacian, corresponds to the kinetic energy, and $V(r)$ corresponds
to the potential energy. Here, we consider a central potential for which the ground state may be expressed as $\psi_{{\rm GS}}\left(\boldsymbol{r}\right) = R_{\rm GS}\left(r\right)/\sqrt{2\pi}$. Inserting Eqs. (\ref{eqpsi_alpha_Bessel})
and (\ref{eqH_0 general}) into Eq. (\ref{eqJ_fiunctional}) one
finds
\begin{align}
\mathcal{J} & =\sum_{n=1}^{N}\sum_{l=\pm}c_{ln}^{\alpha}\left(c_{ln}^{\alpha}\right)^{*}\left[\frac{1}{2\mu}\left(\frac{z_{ln}}{R}\right)^{2}-E_{0}+\omega\right]\nonumber \\
 & +\sum_{n=1}^{N}\sum_{k=1}^{N}\sum_{l=\pm}c_{ln}^{\alpha}\left(c_{lk}^{\alpha}\right)^{*}\mathcal{V}_{kn}^{(l)}\nonumber \\
 & +\frac{1}{2}\sum_{n=1}^{N}\sum_{l=\pm}\left[\left(\delta_{\alpha,x}-il\delta_{\alpha,y}\right)\left(c_{ln}^{\alpha}\right)^{*}\mathcal{S}_{n}^{(l)}+c.c.\right],
\end{align}
where $c.c.$ stands for complex conjugated and the following integrals
where introduced
\begin{align}
\mathcal{V}_{kn}^{(l)} & =\int_{0}^{R}j_{lk}(r)V(r)j_{ln}(r)rdr,\label{eqV matrix el}\\
\mathcal{S}_{n}^{(l)} & =\int_{0}^{R}j_{ln}(r)R_{\rm GS}(r)r^{2}dr.
\end{align}
The first one corresponds to the matrix elements of the potential
between different basis functions, while the second one is proportional
to dipole transitions between the ground state of the unperturbed
system and the functions of the basis. Furthermore,
we note that $\mathcal{V}_{kn}^{(l)}$ is symmetric, that is, $\mathcal{V}_{kn}^{(l)}=\mathcal{V}_{nk}^{(l)}$.
We have omitted the argument of the coefficients $c_{ln}^{\alpha}$
to simplify the notation, however one should keep in mind that these
are $\omega-$dependent quantities.

Now, differentiating $\mathcal{J}$ with respect to the coefficients
$\left(c_{ln}^{\alpha}\right)^{*}$, we obtain a linear system of equations whose solution
determines the coefficients themselves. In matrix notation the linear
system reads
\begin{align}
\mathbb{M}^{(l)}(\omega)\mathbf{c}_{l}^{\alpha}(\omega) & =-\frac{1}{2}\left(\delta_{\alpha,x}-il\delta_{\alpha,y}\right)\mathbf{S}^{(l)},\quad l=\pm1
\end{align}
where
\begin{equation}
\left[\mathbb{M}^{(l)}(\omega)\right]_{ij}=\delta_{ij}g_{j}^{(l)}(\omega)+\mathcal{V}_{ij}^{(l)},
\end{equation}
with 
\begin{equation}
g_{j}^{(l)}(\omega)=\frac{z_{lj}^{2}}{2\mu R^{2}}-E_{0}+\omega,\label{eqg_j(w)}
\end{equation}
and
\begin{align*}
\left[\mathbf{S}^{(l)}\right]^{{\rm T}} & =\left[\mathcal{S}_{1}^{(l)},\mathcal{S}_{2}^{(l)},\ldots,\mathcal{S}_{N}^{(l)}\right],\\
\left[\mathbf{c}_{l}^{\alpha}(\omega)\right]^{{\rm T}} & =\left[c_{l1}^{\alpha}(\omega),c_{l2}^{\alpha}(\omega),\ldots,c_{lN}^{\alpha}(\omega)\right].
\end{align*}
Let us emphasize that to obtain the coefficients that define $|\psi_{\alpha}\rangle$
we need only compute the vector $\mathbf{S}^{(l)}$ and the matrix
$\mathbb{M}^{(l)}$. The most expensive part of the numerical computation
is the calculation of all the $\mathcal{V}_{ij}^{(l)}$. However,
since $\mathcal{V}_{ij}^{(l)}$ is independent of $\omega$ this only
needs to be computed once, regardless of the value of $\omega$ one
wishes to use. The fact that $\mathcal{V}_{ij}^{(l)}$ is symmetric
also greatly reduces the number of integrals that need to be evaluated.
Finally, we point out that when $\alpha=x$ we have $c_{+,n}^{x}=-c_{-,n}^{x},$
since $\mathbb{M}^{(+)}=\mathbb{M}^{(-)}$ and $\mathbf{S}^{(+)}=-\mathbf{S}^{(-)}.$
Following the same reasoning, when dealing with the $y$ direction
we have $c_{+,n}^{y}=c_{-,n}^{y}$, due to the term $il\delta_{\alpha,y}$
which changes sign when $l$ changes sign.

With the problem associated with the functional $\mathcal{J}$ taken
care of, let us move on to the functional $\mathcal{K}.$ Once again
we choose to work in a Fourier-Bessel basis. Now, let us recall that
in the beginning, following Eq. (\ref{eqxi_alpha_beta}), we assumed
$\langle0|\xi_{\alpha\beta}(\omega_{a},\omega_{b})\rangle=0$. In
order to satisfy this, we must have
\begin{align}
\sum_{n=1}^{N}\zeta_{0n}^{\alpha\beta}(\omega_{a},\omega_{b})\int_{0}^{R}j_{0n}(r)R_{{\rm GS}}(r)rdr & =0,
\end{align}
where all the reaming terms in the definition of $\xi_{\alpha\beta}$
are guaranteed to vanish from the angular integration, since for an
isotropic system we have an isotropic ground-state wave function.
This condition can be put in the equivalent form
\begin{align}
\zeta_{01}^{\alpha\beta}(\omega_{a},\omega_{b}) & =-\sum_{n=2}^{N}\zeta_{0n}^{\alpha\beta}(\omega_{a},\omega_{b})f_{n},\label{eqzeta_01}
\end{align}
where
\[
f_{n}=\frac{\int j_{0n}\left(r\right)R_{{\rm GS}}(r)rdr}{\int j_{01}\left(r\right)R_{{\rm GS}}(r)rdr}.
\]
Thus, hereinafter, we no longer consider $\zeta_{01}^{\alpha\beta}$
as an independent variable, but rather as a parameter defined from
the remaining $\zeta_{0n}^{\alpha\beta}$. Inserting Eq. (\ref{eqxi_alpha_beta_Bessel})
in Eq. (\ref{eqK_functional}), and once again using the definition
for $H_{0}$ given in Eq. (\ref{eqH_0 general}), one finds after
some algebra
\begin{align}
\mathcal{K} & =\sum_{n=1}^{N}\sum_{l=-\infty}^{\infty}\zeta_{ln}^{\alpha\beta}\left[\zeta_{ln}^{\alpha\beta}\right]^{*}g_{n}^{(l)}(\omega_{a}+\omega_{b})\nonumber \\
 & +\sum_{n,m=1}^{N}\sum_{l=-\infty}^{\infty}\zeta_{lm}^{\alpha\beta}\left[\zeta_{ln}^{\alpha\beta}\right]^{*}\mathcal{V}_{nm}^{(l)}\nonumber \\
 & +\frac{1}{2}\sum_{n,m=1}^{N}\sum_{s=\pm}\bigg\{ c_{sm}^{\alpha}\left[\zeta_{0n}^{\alpha\beta}\right]^{*}\mathcal{T}_{nm}^{(0,s)}\left(\delta_{\beta,x}+is\delta_{\beta,y}\right)\nonumber \\
 & +c_{sm}^{\alpha}\left[\zeta_{s2,n}^{\alpha\beta}\right]^{*}\mathcal{T}_{nm}^{(s2,s)}\left(\delta_{\beta,x}-is\delta_{\beta,y}\right)+c.c.\bigg\},\label{eqK functional new}
\end{align}
where $c.c.$ stands for complex conjugated, $g_{n}^{(l)}$ and $\mathcal{V}_{nk}^{(l)}$
were defined in Eq. (\ref{eqg_j(w)}) and (\ref{eqV matrix el}),
respectively, and we introduced
\begin{equation}
\mathcal{T}_{nm}^{(l,s)}=\int_{0}^{R}j_{ln}(r)j_{sm}(r)r^{2}dr,\label{eqT_integral}
\end{equation}
which is associated with the dipole transition amplitude between the
functions of the basis. This integral has an analytical solution given
by
\begin{align}
\int_{0}^{1}J_{\nu}(\alpha r)J_{\nu+1}(\beta r)r^{2}dr & =\frac{\alpha J_{\nu+1}(\alpha)}{(\alpha^{2}-\beta^{2})^{2}}\Big[-2\beta J_{\nu}(\beta)\nonumber \\
 & +(\alpha^{2}-\beta^{2})J_{\nu+1}(\beta)\Big],
\end{align}
for any $\nu$ given that $J_{\nu}(\alpha)=0$. When $\beta$ is such
that $J_{\nu+1}(\beta)=0$ (which is our case) the last term vanishes.
From Eq. (\ref{eqT_integral}), we conclude that $\mathcal{T}_{nk}^{(l,s)}$
is not symmetric, since $\mathcal{T}_{nk}^{(l,s)}\neq\mathcal{T}_{kn}^{(l,s)}$.
Since these integrals have analytical solutions, the lack of symmetry
does not significantly impact the numerical efficiency of our approach.
Once again, to simplify the notation, we have omitted the arguments
of the coefficients $c_{lm}^{\alpha}$ and $\zeta_{ln}^{\alpha\beta}$.

With the functional $\mathcal{K}$ in its current form we can differentiate
it with respect to the $\zeta_{ln}^{\alpha\beta}$ and obtain a linear
system in a similar fashion to what was previously done for the functional
$\mathcal{J}$. However, we should remember that in order to satisfy
the relation $\langle0|\xi_{\alpha\beta}(\omega_{a},\omega_{b})\rangle=0$
the coefficient $\zeta_{01}^{\alpha\beta}$ must be treated with care,
since according to Eq. (\ref{eqzeta_01}) it is a function of the
remaining $\zeta_{0n}^{\alpha\beta}$. Thus, it is convenient to deal
with the cases where $l=0$ and $l\neq0$ separately.

Starting with the $l=0$ case, we substitute $\zeta_{0n}^{\alpha\beta}$
in Eq. (\ref{eqK functional new}) by its definition, given in Eq.
(\ref{eqzeta_01}), and differentiate the result with respect to
the $\left(\zeta_{0n}^{\alpha\beta}\right)^{*}$, with $n\geq2$. Proceeding as described
one finds the following linear system defining the coefficients $\zeta_{0n}^{\alpha\beta}$
with $n\geq2$
\begin{equation}
\left[\mathbb{F}+\mathbb{M}^{(0)}(\omega_{a}+\omega_{b})\right]\cdot\boldsymbol{\zeta}_{0}^{\alpha\beta}(\omega_{a},\omega_{b})=-\mathbf{W}_{0}^{\alpha\beta}(\omega_{a})+\mathbf{f}_{0}^{\alpha\beta}(\omega_{a}),\label{eql=00003D0 system}
\end{equation}
where $\mathbb{M}^{(0)}(\omega_{a}+\omega_{b})$ is defined as before,
and
\begin{align}
 & \left(\mathbb{F}\right)_{ij}=\left[g_{1}^{(0)}(\omega_{a}+\omega_{b})+\mathcal{V}_{11}^{(0)}\right]f_{i}f_{j}-\mathcal{V}_{i1}^{(0)}f_{j}-f_{i}\mathcal{V}_{1j}^{(0)},\\
 & \mathbf{W}_{0}^{\alpha\beta}=\frac{1}{2}\sum_{s=\pm}\left(\delta_{\beta,x}+is\delta_{\beta,y}\right)\mathbb{T}^{(0,s)}\cdot\mathbf{c}_{s}^{\alpha},\\
 & \left(\mathbf{f}_{0}^{\alpha\beta}\right)_{n}=\frac{1}{2}f_{n}\sum_{m=1}^{N}\sum_{s=\pm}c_{sm}^{\alpha}\mathcal{T}_{1m}^{(0,s)}\left(\delta_{\beta,x}+is\delta_{\beta,y}\right),
\end{align}
with $\left(\mathbb{T}^{(0,s)}\right)_{ij}=\mathcal{T}_{ij}^{(0,s)}$,
and
\[
\left[\boldsymbol{\zeta}_{0}^{\alpha\beta}\right]^{{\rm T}}=\left[\zeta_{02}^{\alpha\beta},\zeta_{03}^{\alpha\beta},\ldots,\zeta_{0N}^{\alpha\beta}\right].
\]
We note that the vectors $\boldsymbol{\zeta}_{0}^{\alpha\beta}$,
$\mathbf{W}_{0}^{\alpha\beta}$ and $\mathbf{f}_{0}^{\alpha\beta}$
are $(N-1)\times1$; the vector $\mathbf{c}_{s}^{\alpha}$ is $N\times1$;
the matrices $\mathbb{F}$ and $\mathbb{M}^{(0)}$ are $(N-1)\times(N-1)$
and the matrix $\mathbb{T}^{(0,s)}$ is $(N-1)\times N$. The solution
of this system gives the $\zeta_{0n}^{\alpha\beta}$ with $n\geq2$,
from which the value of $\zeta_{01}^{\alpha\beta}$ can be computed.

Having dealt with the delicate case of $l=0$ we can now study the
contributions originating from the cases where $l\neq0$. Since no
restrictions are imposed on coefficients with $l\neq0$ this is a
simpler problem. Returning to Eq. (\ref{eqK functional new}), and
differentiating $\mathcal{K}$ with respect to the $\left(\zeta_{ln}^{\alpha\beta}\right)^{*}$,
with $n\geq1$ and $l\neq0$, one finds
\begin{equation}
\mathbb{M}^{(l)}(\omega_{a}+\omega_{b})\cdot\boldsymbol{\zeta}_{l}^{\alpha\beta}(\omega_{a},\omega_{b})=-\mathbf{W}_{l}^{\alpha\beta}(\omega_{a}),\quad l\neq0,\label{eql neq 0 system}
\end{equation}
where $\mathbb{M}^{(l)}$ and $\boldsymbol{\zeta}_{l}^{\alpha\beta}$
are defined as before, only this time they are $N\times N$ and $N\times1$,
respectively. The definition of $\mathbf{W}_{l}^{\alpha\beta}(\omega_{a})$
reads 
\begin{align}
\mathbf{W}_{l}^{\alpha\beta}(\omega_{a}) & =\frac{1}{2}\delta_{l,2}\left(\delta_{\beta,x}-i\delta_{\beta,y}\right)\mathbb{T}^{(2,1)}\cdot\mathbf{c}_{+}^{\alpha}(\omega_{a})^{{\rm }}\nonumber \\
 & -\frac{1}{2}\delta_{l,-2}\left(\delta_{\beta,x}+i\delta_{\beta,y}\right)\mathbb{T}^{(2,1)}\cdot\mathbf{c}_{-}^{\alpha}(\omega_{a})^{{\rm }}.\quad l\neq0
\end{align}
This linear system is numerically well behaved and, therefore, can
be solved with any linear-algebra numerical package. Its solution
gives the coefficients $\boldsymbol{\zeta}_{l}^{\alpha\beta}$, with
$l\neq0$, necessary to compute $|\xi_{\alpha\beta}\rangle$. Comparing
Eq. (\ref{eql neq 0 system}) with Eq. (\ref{eql=00003D0 system}),
we observe that their structure is very much alike, the only difference
being the appearance of $\mathbf{f}_{0}^{\alpha\beta}$ and $\mathbb{F}$
in Eq. (\ref{eql=00003D0 system}). These two terms have their origin
on the restriction imposed by the condition $\langle0|\xi_{\alpha\beta}(\omega_{a},\omega_{b})\rangle=0$,
and thus do not appear in Eq. (\ref{eql neq 0 system}). For both
the cases where $l=0$ and $l\neq0$, it is necessary to first solve
the problem associated with the functional $\mathcal{J}$ in order
to obtain the coefficients $\mathbf{c}_{l}^{\alpha}(\omega_{a})$.
Moreover, it is clear that the terms with $l=\pm2$ play a distinct
role in the problem. In fact, the only relevant terms are the ones
with $l=0,\pm2$, since only they yield finite matrix elements when
the susceptibility is computed. Terms with different values of $l$
vanish when the angular part of the matrix elements is calculated.
Finally, we note that since $\mathbb{M}^{(2)}=\mathbb{M}^{(-2)}$
and $\mathbf{W}_{2}^{\alpha\beta}=\mathbf{W}_{-2}^{\alpha\beta}$
when $\alpha=\beta$, we have $\boldsymbol{\zeta}_{2}^{\alpha\beta}=\boldsymbol{\zeta}_{-2}^{\alpha\beta}$
when $\alpha=\beta$. If $\alpha\neq\beta$, then $\boldsymbol{\zeta}_{2}^{\alpha\beta}=-\boldsymbol{\zeta}_{-2}^{\alpha\beta}$.

\section{Details on the circular well problem\label{secDetails-on-the}}

In this appendix we give a detailed description of the necessary calculations
to compute the TPA of the circular well. We start by writing the wave
function $\psi_{x}(\omega,\mathbf{r})$ as
\begin{equation}
\psi_{x}(\omega,\mathbf{r})=\sqrt{\frac{2}{\pi}}\sum_{n=1}^{N}c_{+,n}^{x}(\omega)j_{1n}(r)\cos\theta,\label{eqpsi_x WF}
\end{equation}
where we used the fact that $c_{+,n}^{x}=-c_{-,n}^{x}$ (see Appendix
\ref{secComputing-the-new}). Regarding the wave function $\xi_{xx}(\omega_{1},\omega_{2};\mathbf{r})$,
and using Eq. (\ref{eqxi_alpha_beta_Bessel}), we obtain
\begin{align}
\xi_{xx}(\omega_{1},\omega_{2};\mathbf{r}) & =\frac{1}{\sqrt{2\pi}}\sum_{n=1}^{N}\bigg\{\zeta_{0n}^{xx}(\omega_{1},\omega_{2})j_{0n}(r)\nonumber \\
 & +2\zeta_{2n}^{xx}(\omega_{1},\omega_{2})j_{2n}(r)\cos2\theta\bigg\},\label{eqxi_xx WF}
\end{align}
where the relation $\zeta_{2n}^{xx}=\zeta_{-2n}^{xx}$ was used (see
Appendix \ref{secComputing-the-new}). To obtain $\chi_{xxxx}^{{\rm TPA}}(\omega)$
we have to compute three different types of matrix elements, which
can be written in a fairly compact form using Eq. (\ref{eqpsi_x WF})
and (\ref{eqxi_xx WF})
\begin{align}
 & \langle\psi_{x}(\omega_{2}^{*})|\psi_{x}(\omega_{1})\rangle=\sum_{n=1}^{N}c_{+,n}^{x}(\omega_{2}^{*})^{*}c_{+,n}^{x}(\omega_{1})\\
 & \langle0|\mathbf{d}_{x}|\psi_{x}(\omega_{1})\rangle=\mathbf{c}_{+}^{x}(\omega_{1})\cdot\mathbf{S}^{(+)}\\
 & \langle\psi_{x}(\omega_{1}^{*})|\mathbf{d}_{x}|\xi_{xx}(\omega_{2},\omega_{3})\rangle=\bigg(\left[\boldsymbol{\zeta}_{0}^{xx}(\omega_{2},\omega_{3})\right]^{{\rm T}}\cdot\mathbb{T}^{(0,1)}\nonumber \\
 & +\left[\boldsymbol{\zeta}_{2}^{xx}(\omega_{2},\omega_{3})\right]^{{\rm T}}\cdot\mathbb{T}^{(2,1)}\bigg)\cdot\mathbf{c}_{+}^{x}(\omega_{1}^{*})^{*},
\end{align}
where the vector $\mathbf{S}^{(+)}$ and the matrices $\mathbb{T}$
were first introduced when the functionals $\mathcal{J}$ and $\mathcal{K}$
were studied. The fact that these only need to be computed once, but
appear in different instances of the calculation contributes to the
simplicity and efficiency of the approach.

The only thing left to do is to compute all the necessary coefficients
$\mathbf{c}_{+}^{x}$, $\boldsymbol{\zeta}_{0n}^{xx}$ and $\boldsymbol{\zeta}_{2n}^{xx}$.
Since inside the disk where the problem is defined the potential vanishes,
all the terms containing $\mathcal{V}_{nk}^{(l)}$ disappear; this
significantly simplifies the computation of the coefficients. The
$\mathbf{c}_{+}^{x}$ are given by
\begin{equation}
\left[\mathbf{c}_{+}^{x}(\omega)\right]_{j}=-\frac{1}{2g_{j}^{(+)}(\omega)}\left[\mathbf{S}^{(+)}\right]_{j},\quad1\leq j\leq N.
\end{equation}
It is easily verified that these coefficients quickly approach zero
even for modest values of $N$. This is a direct consequence of the
fast decay of $\mathcal{S}_{j}^{(+)}$ as $j$ increases. To compute
the $\boldsymbol{\zeta}_{0n}^{xx}$ the first thing to note is that
for the circular disk, where the ground state wave function is proportional
to the Bessel function $J_{0}(z_{01}r/R)$, all the $f_{j}$ vanish,
due to the orthogonality relation of Bessel functions on a disk. As
a consequence, $\zeta_{01}^{xx}=0$. The remaining $\boldsymbol{\zeta}_{0n}^{xx}$
follow from
\begin{equation}
\boldsymbol{\zeta}_{0}^{xx}(\omega_{a},\omega_{b})=-\left[\mathbb{M}^{(0)}(\omega_{a}+\omega_{b})\right]^{-1}\cdot\mathbb{T}^{(0,1)}\cdot\mathbf{c}_{+}^{x}(\omega_{a}),
\end{equation}
where $\boldsymbol{\zeta}_{0}^{xx}$ is a $(N-1)\times1$ vector.
This becomes a $N\times1$ vector once the value of $\zeta_{01}^{xx}=0$
is introduced. The inverse of the matrix $\mathbb{M}^{(0)}$ is simply
given by $\left[\mathbb{M}^{(0)}\right]_{j}^{-1}=1/g_{j}^{(0)}$.
Finally, to compute the $\boldsymbol{\zeta}_{2n}^{xx}$ one uses
\begin{equation}
\boldsymbol{\zeta}_{2}^{xx}(\omega_{a},\omega_{b})=-\frac{1}{2}\left[\mathbb{M}^{(2)}(\omega_{a}+\omega_{b})\right]^{-1}\cdot\mathbb{T}^{(2,1)}\cdot\mathbf{c}_{+}^{x}(\omega_{a}),
\end{equation}
where $\left[\mathbb{M}^{(2)}\right]_{j}^{-1}=1/g_{j}^{(2)}$. The
fast convergence of the $\mathbf{c}_{+}^{x}$ aides the convergence
of the various $\boldsymbol{\zeta}^{xx}$ coefficients.


\end{document}